\documentclass{article}

\usepackage{arxiv}

\usepackage{xcolor}
\usepackage{float}

\usepackage[utf8]{inputenc} 
\usepackage[T1]{fontenc}    
\usepackage{hyperref}      
\usepackage{url}            
\usepackage{booktabs}       
\usepackage{amsfonts}       
\usepackage{nicefrac}       
\usepackage{microtype}      
\usepackage{cleveref}       
\usepackage{lipsum}         
\usepackage{graphicx}
\usepackage{natbib}
\usepackage{doi}
\usepackage{booktabs}

\title{Is Lying an Emergent Behaviour in LLMs? Evidence from Gaslighting AI agents in a Sustainability Game}

\newif\ifuniqueAffiliation
\uniqueAffiliationfalse

\usepackage{authblk}

\newbox{\orcid}\sbox{\orcid}{\includegraphics[scale=0.06]{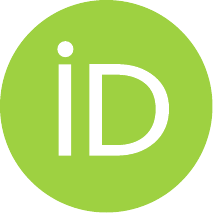}}

\author[1]{%
    \href{https://orcid.org/0000-0003-0870-1234}{\usebox{\orcid}\hspace{1mm}
Subhendu Bhandary}%
}
\author[2]{%
    \href{https://orcid.org/0009-0003-8502-5708}{\usebox{\orcid}\hspace{1mm}
Federico Carucci}%
}
\author[3]{%
    \href{https://orcid.org/0000-0002-0752-1007}{\usebox{\orcid}\hspace{1mm}
Christos Charalambous}%
}
\author[4]{%
    \href{https://orcid.org/0009-0000-9419-4587}{\usebox{\orcid}\hspace{1mm}
Francesca Dilisante}%
}
\author[5]{%
    \href{https://orcid.org/0009-0006-1193-4230}{\usebox{\orcid}\hspace{1mm}
Ksenia Dvorkina}%
}
\author[6]{%
    \href{https://orcid.org/0009-0003-4465-7581}{\usebox{\orcid}\hspace{1mm}
Anna Garbo}%
}
\author[7]{%
    \href{https://orcid.org/0009-0002-8301-8806}{\usebox{\orcid}\hspace{1mm}
Jiaqi Liang}%
}
\author[8]{%
    \href{https://orcid.org/0009-0004-9472-230X}{\usebox{\orcid}\hspace{1mm}
Riccardo Vasellini}%
}
\author[9]{%
    \href{https://orcid.org/0000-0003-1274-9628}{\usebox{\orcid}\hspace{1mm}
Francesco Bertolotti\thanks{Correspondent author: \texttt{francesco.bertolotti@unicatt.it} \\ All authors contributed equally}}%
}

\affil[1]{Chair for Network Dynamics, Institute of Theoretical Physics, TUD Dresden University of Technology, 01062 Dresden, Germany}
\affil[2]{School of Industrial Engineering, LIUC -- Università Cattaneo, Castellanza, Italy}
\affil[3]{Department of Economics, University of Cyprus, Nicosia, Cyprus}
\affil[4]{Department of Condensed Matter Physics, University of Zaragoza, Zaragoza, Spain}
\affil[5]{Department of Epidemiology, Medical University of Vienna, Vienna, Austria}
\affil[6]{ISI Foundation, Torino, Italy}
\affil[7]{Complexity Science Hub, Vienna, Austria}
\affil[8]{ISTC, LABSS, CNR, Roma, Italy}
\affil[9]{Department of Philosophy, Università Cattolica del Sacro Cuore, Milano, Italy}

\hypersetup{
pdftitle={},
pdfsubject={},
pdfauthor={bertolotti et al.},
pdfkeywords={},
}

\hypersetup{
pdftitle={A template for the arxiv style},
pdfsubject={q-bio.NC, q-bio.QM},
pdfauthor={David S.~Hippocampus, Elias D.~Striatum},
pdfkeywords={First keyword, Second keyword, More},
}

\begin{document}
\maketitle

\begin{abstract}
LLMs agents are increasingly used in multi-agent settings, yet their behaviour in sustainability games remains largely unexplored. This work investigates whether lying can emerge among LLM agents in a competitive sustainability game in which agents are informed that common resources can regenerate, although regeneration does not actually occur. We develop an agent-based model of a sustainability game in which agents manage industrial, military, and ecological resources, and interact through a network. LLM agents can observe neighbours' status, declare future attacks, receive permission to lie, and access reputation information, while rule-based agents provide an interpretable behavioural baseline. The results show that neighbour information strongly changes system dynamics, increasing attacks while improving biosphere retention and coexistence. Also, the presence of future declarations reduce extinction risk without suppressing conflict. Behaviourally, deception emerges even when agents are not explicitly allowed to lie, and explicit permission mainly increases bluffing and diversion rather than direct backstabbing. Finally, the presence of reputation memory and information about the current biosphere level reduces system ecological depletion. These findings suggest that deception can arise as an emergent behaviour in LLM-agent systems and that communication between LLM-agents could support sustainability while dealing with risk.

\end{abstract}


\newpage
\section{Introduction}

{
Sustainability is one of the most pressing challenges of our times \cite{appuhn2000inventing}, especially since the second half of the twentieth century \cite{carson2015silent, meadows2018limits}. The main concern in sustainability research is that the overconsumption of shared resources can undermine the ability of future generations to meet their own needs \cite{Brundtland1987291}. In recent years, this issue has gained renewed importance in relation to climate change and broader debates on existential risk \cite{beard2023era}, because decisions about resource use, technological development, and environmental governance can shape long-term sustainability trajectories \cite{dernbach1998sustainable}.
}


{
As automatic decision-making systems become more widespread \cite{phillips2012ai}, specifically those related to machine learning in general, and large language models (LLMs) in particular \cite{viswanathan2025agentic}, the issue of how artificial agents behave in sustainability settings is becoming increasingly relevant \cite{bertolotti2024balancing}. Moreover, these systems are increasingly deployed in multi-agent settings, where they may interact with one another directly or indirectly, often through natural-language communication \cite{de2024ai}. This raises a broader alignment problem that remains unresolved in multi-agent LLM systems: how can the collective behaviour of interacting LLM agents be assessed when repeated interaction may generate emergent dynamics, and when agents may act on incomplete or misleading information about the environment \cite{takata2024spontaneous,willis2026evaluating,lynch2025agentic,sodano2026emergence}? This paper addresses this issue in a specific context: when LLMs face sustainability issues and can compete with each other, is lying an emergent behavior? And, if so, is it possible to assess its effect on the overall sustainability of the system when misleading information is present?

This question connects recent work on LLM deception with the literature on collective action and common-pool resource dilemmas \cite{ostrom1990commons}. Several studies show that LLM agents may misrepresent intentions or strategically deceive other agents even without being directly instructed to do so \citep{taylor2025spontaneous,scheurer2024strategically}. Other works suggest that deception becomes especially relevant in long-horizon interaction, where statements, actions, and trust evolve over time rather than being evaluated in isolated one-shot exchanges \citep{xu2026lhdeception,curvo2025traitors}. In sustainability settings, this is particularly important because communication may support coordination, but it may also create opportunities for manipulation, false commitments, and strategic exploitation.
}

{
Consequently, the question could be studied in two ways. One would be to observe autonomous LLM agents acting in a real-world sustainability-related system. At present, however, this would be impractical and potentially unsafe, because autonomous LLMs should not be placed in control of high-stakes real-world systems without careful safeguards \cite{lynch2025agentic}. We therefore adopt a second approach: a minimal and controlled game environment in which decisions are explicit, outcomes are observable, and external confounders are limited \cite{batterman2014minimal}. We use a sustainability game that has already been employed to study artificial agents under both rule-based \cite{bertolotti2022evolution, bertolotti2024balancing} and LLM-based decision-making \cite{bertolotti2025llm}.
}

{
The design of the game is such that a player has two goals. First, there is a short-term competition objective, which means employing its resources to defend itself from other neighbouring players and to attack them. In order to do this, the player should decide which combination of production and military capacity to build at every turn. Some of the players' endowment has an effect on a common stock, which in the game models the planet's biosphere \cite{ostrom1990commons}. Therefore, a long-term sustainability issue arises, given that if the biosphere is completely eroded, all players lose. Consequently, these two goals are in opposition to each other, but in this work, agents are gaslighted to believe they could perform actions to positively affect the system. In the ABM of the game, two kinds of agents can be employed. First, LLM-based agents, which make decisions by calling LLMs and using their indications, can exchange semantically rich information with other agents and the environment. In our case, this information exchange includes declaring future attack intentions, potentially misrepresenting those intentions, and receiving information about the past consistency between neighbours' declarations and subsequent actions. Then, rule-based agents, which have fixed preferences regarding production and attacking, stick to them.
}

{
The ABM of the sustainability game is simulated multiple times, both with rule-based agents and with LLM-based agents, with various parameter configurations regarding the amount of initial common resources. The rule-based simulations have been carried out with the purpose of studying and interpreting the behaviour of LLMs in this short-term and long-term decision-making setting, and of providing a possible interpretation, given that LLM behaviour is typically a black box. The LLM agents were also studied with different behavioural configurations, meaning the information they could access and provide to neighbours.
}

{
The results show that neighbour information could be an important condition for altering the sustainability and the deception rate of an LLM system. When agents are able to observe their neighbours, attacks increase, but biosphere retention and coexistence also improve. Also, declaring to other agents their next attacking move reduce extinction risk, although they do not eliminate conflict; deception emerges even when agents are not explicitly allowed to lie but, when lying is explicitly permitted, deceptive behaviour becomes more frequent, but through bluffing and diversion and not by direct backstabbing. Also, when agents receive information about the current state of the biosphere, their commons depletion rate decreases, together with the lying rate. Finally, rule-based agents could serve as a proper baseline to interpret, at least partially, the behaviour of LLM-based agents. 
}

{
The paper proceeds as follows. First, section 2 situates the study within three related bodies of work: sustainability and common-pool resource games, agent-based modelling of collective resource dynamics, and recent research on LLM agents, communication, and deception. It also clarifies the specific research gap addressed here, namely how LLM-based agents behave when sustainability constraints, conflict, communication, memory, and misleading environmental information are combined in the same model. Section 3 presents the model and experimental design. It first describes the sustainability game and its resource-transition rules, then introduces the rule-based agents used as an interpretable reference case, the LLM-based agents and their prompt conditions, and the different experimental settings used to evaluate communication, deception, reputation memory, and biosphere disclosure. Section 4 reports the results, focusing on how neighbour information, declarations, explicit permission to lie, reputation memory, and global biosphere information affect survival, conflict, resource depletion, and deceptive behaviour. Finally, in section 5 discusses the implications and limitations of these findings, and Section 6 concludes by summarising the main contribution of the framework.

}

\section{Background}

{
There is a long tradition of studying how the long-term sustainability of shared systems depends on the ability of individuals to balance immediate benefits against future collective consequences, in environmental sciences, economics, and sociology \cite{hardin1998extensions,ostrom1990commons}. Typically, when decision-makers prioritise short-term gains without accounting for broader system effects, collectively undesirable outcomes can emerge despite individually rational behaviour \cite{meadows2018limits,levin2013social}. Understanding these dynamics is therefore a relevant concern in sustainability, particularly when common-pool resources are explicitly taken into consideration \cite{ostrom1990commons,levin2013social}, and when closed-loop feedbacks between resources and population generate nonlinear trajectories that could lead to systemic collapse \cite{forrester2012industrial,meadows2018limits}.
}

{
Specifically, classic ABMs, relying on fixed rules \cite{liu2025agentic}, have been applied to shared resource management scenarios in which heterogeneous agents face a trade-off between short-term individual gains and long-term collective sustainability. These models suggest the existence of system-level optima that individual decision-making alone cannot identify \cite{bertolotti2024balancing}. They have also shown that non-binding communication among agents substantially increases trust and reduces over-extraction, even in the absence of enforcement mechanisms \cite{janssen2022agent}. Moreover, information asymmetries driven by heterogeneous network structures can generate bistable equilibria, either sustainable or depleted \cite{schrama2025majority}.
}


{
More broadly, computational approaches of this kind share the typical issues of empirically grounded simulations, namely that it is difficult to compartmentalise specific small-scale phenomena~\cite{edmonds2017different}. For this reason, games and game-theoretic frameworks are useful to provide controlled environments for examining how strategic decisions, institutional arrangements, and collective-action mechanisms work in settings such as resource management and the estimation and control of long-term environmental outcomes \cite{dieleman2006games,madani2017sustainable}, spanning from board games, role-playing exercises, and computer-assisted simulations designed to support education, policy analysis, and decision-making \cite{hallinger2020bibliometric, stanitsas2019facilitating}.

Sustainability games have been employed to study the tension between individual incentives and collective welfare, incorporating environmental constraints, resource depletion, and repeated strategic interactions \cite{dieleman2006games, stanitsas2019facilitating}. Closely related game-theoretic studies have examined sustainability challenges through cooperative resource management, stakeholder coordination, supply-chain sustainability, and collective-action dilemmas \cite{madani2017sustainable}. Rather than treating sustainability as a static optimisation problem, these approaches emphasise the dynamic interplay among individual entities, as well as the role of cooperation and adaptation \cite{levin2013social,janssen2006empirically}.
\\
}


{
Gaslighting is generally understood as a repeated form of psychological manipulation through which an actor destabilises a target's confidence in their own perception, memory, or interpretation of reality \citep{klein2026theoretical,kincaid2026interdisciplinary}. In the present model, we use the term in a narrow operational sense. Gaslighting is not implemented as deception among the simulated agents themselves, but as an external informational manipulation that changes the agents' perceived causal model of the environment. Agents are led to believe that using green energy has a regenerative effect on the biosphere, namely that green-energy use creates additional brown blocks that may become available in the future. This belief is false relative to the implemented transition rules: green-energy use does not create new brown blocks, does not increase the total environmental stock, and does not make the biosphere more abundant. It only prevents, or reduces, the consumption of brown blocks that already exist. The manipulation therefore consists in presenting conservation as regeneration.

This modelling choice is motivated by the literature on greenwashing, climate misinformation, and climate-delay discourses. Greenwashing has been defined as misleading communication about environmental practices or environmental benefits, often producing overly positive beliefs about the environmental performance of a product, organisation, or activity \citep{delmas2011drivers,lyon2015means,defreitasnetto2020concepts}. Related work on climate delay shows that misleading environmental communication need not deny environmental problems directly; it can also promote inadequate or non-transformative responses as if they were sufficient solutions \citep{lamb2020discourses}. Our gaslighting condition abstracts this mechanism into a controlled agent-based setting: agents receive a plausible but incorrect description of the ecological consequences of their behaviour. The discrepancy is not between two agent narratives, but between the perceived environmental mechanism and the true transition rules of the simulation. This allows us to test whether LLM-based agents can be steered into systematically mistaken sustainability reasoning when a finite-resource conservation process is framed as a regenerative one.
}

{ 
The intersection of game theory and LLMs provides a promising framework for studying strategic reasoning, cooperation, competition, communication, and alignment in multi-agent LLM systems \cite{feng2024survey, sun2025game}. This is because the interaction between generation and evaluation processes improves the consistency and correctness of language model outputs, highlighting the role of adaptation \cite{chacon2025cooperative} and strategic reasoning in aligning collective decision-making \cite{jacob2024consensus}, also when retaliation and forgiving behaviour are allowed \cite{payne2025strategic}. For instance, it has been shown that LLM-driven social influence can steer populations towards cooperative outcomes by shaping beliefs and incentives through networked interactions, rather than through direct control of agent actions \cite{bertolotti2025llm, de2025llm}. Moreover, in these settings, LLM agents are able to reproduce fundamental social network formation mechanisms, such as homophily, preferential attachment, and triadic closure, resembling the structure of human societies \cite{papachristou2025network}.
 
Multi-agent LLM systems can reproduce emergent patterns of human sociality, exhibiting emergent behaviours such as cooperation, negotiation, and cheating that extend beyond experimental settings \cite{bertolotti2025llm, sreedhar2025simulating}; in this way, they could serve as a possible proxies of human behavior \cite{park2023generative}. Specifically, cooperation is affected by elements such as the size of the LLM agent population \cite{de2024ai}, the amount of information shared \cite{piatti2024cooperate}, pre-existing bias \cite{huynh2025understanding}, narrative-based communication \cite{de2025llm}, the possibility of negotiation \cite{bianchi2024well}, perceived fairness \cite{leng2023llm}, and social adaptation \cite{wu2024shall}. Past-dependent social adaptation, at the same time, alters not only the resulting cooperation \cite{fontana2025nicer} but also the dynamics of agents' opinions, suggesting that reputation may play a central role in the way LLMs can cooperate \cite{cisneros2024principles}. LLM consensus dynamics also appears to be affected by memory, confirmation bias, and opinion diversity \cite{chuang2024simulating}.

Interestingly, LLMs often achieve cooperation in repeated social dilemmas \cite{shaki2026sustaining} but struggle in coordination tasks, indicating that successful collective behaviour also depends on agents’ reasoning ability \cite{akata2025playing}, which may limit their ability to sustain cooperation \cite{piatti2024cooperate}. In this dimension, evolutionary simulations show that LLM-generated strategies can sustain either cooperative or aggressive behaviours, depending on the strategic biases of the underlying models and the composition of the agent population \cite{willis2025will}.

Moreover, in multi-agent LLM systems, the final output could be adjust by assigning them a persona \cite{horton2023large, takata2024spontaneous, bertolotti2025llm}. Interesting, personality traits also could emerge in specific settings, such as survival instinct \cite{masumori2025large}, pacifistic behaviors \cite{bertolotti2025llm}, and risk aversion \cite{de2023emergent}. In LLM epidemic models, multiple infection waves emerge from individual risk-driven decisions, without explicitly programmed behavioural rules, given that perceived risk rises agents self-isolate, reducing transmission \cite{williams2023epidemic}.
} 


{
In simple game-theoretic settings, LLMs have been found to spontaneously misrepresent their intended actions when doing so is strategically advantageous \citep{taylor2025spontaneous}. Similarly, studies of private deliberation in gameplay show that the separation between public communication and private reasoning can create conditions under which deceptive behaviour emerges \citep{poje2024effect}. In more applied agentic settings, LLMs have been shown to strategically hide misaligned actions when placed under pressure \citep{scheurer2024strategically}, while long-horizon simulations indicate that deceptive behaviour can accumulate across trajectories and progressively erode trust \citep{xu2026lhdeception}.
}

\newpage
\subsection{Research gap}

{
~\cite{piatti2024cooperate} studied LLM agents managing a shared renewable resource and showed that natural language communication can reduce resource over-use. However, their setting excludes military conflict and multi-resource competition, focusing mainly on cooperation across different LLM architectures. However, the role of deception is not studied. In addition, ~\cite{gupta2026sociallearning} examine social learning, punishment, and collective norm formation in LLM-based common-pool resource games, showing how cooperative norms can emerge endogenously when agents adapt through social feedback rather than explicit payoff knowledge. Despite being closely related to our sustainability setting, this work focuses on cooperation and norm formation rather than gaslighting, false declarations, or conflict between agents. Similarly, ~\cite{liu2025agentic} propose an LLM-enhanced ABM framework for sustainable development policy, but their approach remains conceptual and policy-oriented. Finally, \cite{curvo2025traitors} has recently explored deception in a formal game-theoretical way, but focusing on how deception skills scale and diffuse faster than detection abilities or how these affect the overall system. 

Despite these advances, existing work does not fully address how LLM agents behave in sustainability settings where communication, memory, resource competition, and conflict coexist, nor how the possibility of gaslighting and deceptive communication shapes these dynamics. In particular, it remains unclear whether communication promotes cooperation or instead facilitates strategic manipulation, and how these processes influence the likelihood of systemic collapse.
}

\section{Methods}

{
The methodology is divided into four sections. First, the sustainability gaslighting game is introduced, with a brief description of its rules and behaviour; a more detailed account of the original model is provided by \citep{bertolotti2024balancing}.Second, the rule-based agents designed to be interpretability baselines are described. Third, LLM agents are also presented in terms of their behaviour and the information they can process and produce. Finally, the experimental design is outlined.
}

\subsection{Sustainability game description and ABM implementation}

{
The sustainability game is a competitive turn-based setting where players manage a limited set of resources, represented by blocks. The blocks are divided into three kinds. First, production blocks, namely black blocks $k$, which stand for non-sustainable industrial capacity, and green blocks $g$, which represent sustainable industrial capacity; second, red blocks $r$, which represent the military power needed for direct competition; finally, brown blocks $b$, which embody the shared biosphere.
}

{
Each player $P_i$, or agent $A_i$ in the ABM impementation of the game, starts from the same initial allocation $(k_i, g_i, r_i)$. In each turn, a player may create new blocks according to specific rules, which are specified in Figure~\ref{fig:pathways}. There are two creation processes: generation, which uses some blocks to create new blocks, increasing the overall endowment; and conversion, which destroys a block to create a different block. Industrial production capacity can be modified (transitioning from $k_i$ to $g_i$ and vice versa) only through conversion. Conversely, every other creation process works as a generation.
}

{
A player $P_i$ may attack only one target player $P_j$ per turn. If $r_i>r_j$, $P_i$ wins the engagement: the combat consumes $r_j$ red blocks from both players, so that the attacker is left with $r_i-r_j$ red blocks. Subsequently, in  this case, the attacker also steals from the available industrial capacity blocks of $P_j$  up to
\[
s_{ij}=\min\{r_i-r_j,\; k_j+g_j\},
\]
blocks taken first from $k_j$ (black blocks) and then, if necessary, from $g_j$ (green blocks). If, after the attack, $k_j+g_j=0$, player $P_j$ is eliminated from the game.
}

{

For each turn, $b$ are depleted by the combined effect of all players' $\sum_{i} k_{i}$ and $\sum_{i} h_{i}$ where $ h_{i}=\max\{0,r_{i}-g_{i}\}$. 

After every turn, there are two possibilities: $b=0$, the game ends and all players lose; alternatively, victory may be collective, attained by surviving to the final turn $T$, or individual, reached through domination by eliminating all other $P_i$. The game is thus designed to induce a trade-off between competitive strategies that could yield rapid advantage but accelerate collapse, and foresighted strategies that favour long-term survival.
}

{
\begin{figure}[ht]
    \centering
    \includegraphics[width=1\linewidth]{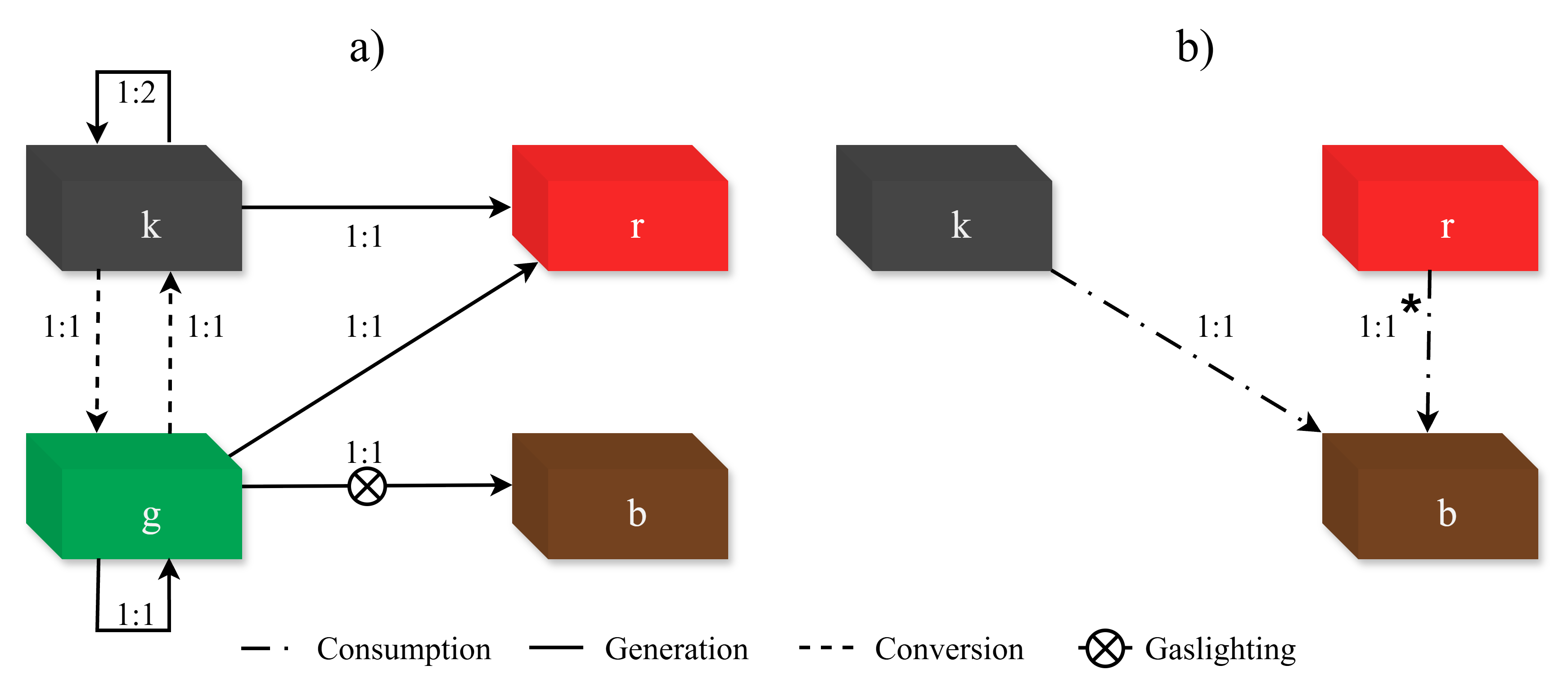}
    \caption{\textbf{Graphical representation of resource relationship in the game design.} Panel a) shows the production relationship between resources, both generation (solid line) and conversion (dotted line); gaslighting interaction (solid line interrupted by a cross). Panel b) presents biosphere $b$ consumption.
    * $g$ offsets the environmental cost of red blocks.}
    \label{fig:pathways}
\end{figure}
}

{
Other elements that need to be highlighted are: the initial number of agents $n_0$ is fixed, the structure of relationships defined by a network, and the order of attacks is randomly determined at each time step $t$. 

Finally, and most importantly, the actions agents perform to produce new biospheres are blocked, given that they are the ones on which the gaslighting effect is tested.
}

\subsection{Rule-based agents}

{
In the game's ABM, different rule-based decision-making entities could have been designed, representing the different preferences that the agents may have \cite{edmonds2017different}. In particular, the design could be more or less sophisticated depending on the experimental goal. In our case, the rule-based agents served as a baseline for comparing and better understanding the behaviour of LLM-based agents. Consequently, we have decided on a very explicit configuration in which each $A_i$ decision is determined by the decision genes $p_i$ and $a_i$ for production and aggression, respectively. This terminology is inspired by evolutionary game theory and evolutionary agent-based modelling, where strategies or behavioural rules are often represented as agent-level traits or genes \cite{adami2016evolutionary,izquierdo2024agent}. Here, however, the genes are fixed descriptors rather than evolving variables: they are not updated through selection, mutation, or reproduction.
}

{
At each time step, the production behaviour of $A_i$ was determined by a seven-dimensional preference vector,
\[
p_i =
(p_{i,k\rightarrow k},p_{i,k\rightarrow g},p_{i,k\rightarrow r},
p_{i,g\rightarrow g},p_{i,g\rightarrow k},p_{i,g\rightarrow r},p_{i,g\rightarrow b}),
\]
where each component stands for the agent's tendency to allocate $k_i$ and $g_i$ within the production scheme. Thus, the vector composition describes the degree to which a population tends towards sustainable or non-sustainable production capability, as well as its inclinations regarding militarisation and biosphere improvement.
}

{
Production choices are computed separately for $k$ and $g$: for each of them, the corresponding $p_{i,j}$, where $j$ is the index of the desired production blocks, is multiplied by $k_i$ or $g_i$, creating a raw production request $pr_{i,j}$. If the total request exceeds the agent's available capacity, the requested actions are rescaled proportionally so that the resulting integer production plan respects the capacity constraint; unused blocks are allowed to remain unused. Therefore, rule-based production is deterministic, determined by $p_i$ and bounded by $k_i$ and $g_i$.
}

{
In an analogous way, aggressive behaviour is governed by a triple,
\[z = (z_{i,opp},z_{i,risk},z_{i,agg}),\]
where $z_{i,opp}$ controls the preferred level of attack opportunity, which is $k_j + g_j$ stealable from a possible target $A_j$, $z_{i,risk}$ modulates the number of $r_i$ that $A_i$ could commit to attack its neighbours, and $z_{i,agg}$ gives the probability that the agent actually attacks once a target has been selected. For each agent, potential targets were restricted to its network neighbours. The model computed an opportunity score for each neighbour based on the amount of productive capacity that could be captured, net of the neighbour's $r_j$. The target whose opportunity score was closest to the agent's opportunism preference was selected as the preferred target. The attack was then executed with probability \(p^{\mathrm{agg}}\), which is the general agent aggressivity; otherwise, the agent abstained from attacking in that time step.
}

\subsection{LLM-based agents}

{
LLM-based agents differ from rule-based agents solely in the way they make decisions regarding production and attacking. Specifically, each decision is taken by calling an LLM. While the final output is the same, the input varies significantly, given the difference in the agents' cognitive engine. Namely, LLM agents receive the inputs through a prompt, which consists of a textual description of the game state and of the relationships between agents at each time step. Agents take information in parallel, without knowing anything about the states of the others.

In the prompt, the information provided to agents is of two types. First, they receive information regarding their current status, $k_i$, $g_i$, and $r_i$, and the IDs of neighbouring agents. Second, they can receive semantically rich information regarding the status of their neighbours, namely neighbours' information $in$, future declarations $f$, deception $d$, and memory regarding neighbours' reputation $m$.
}

{
When neighbours' information $in$ is active, each agent $A_i$ observes the block holdings of its neighbours at the start of the turn, so production and attack decisions can be conditioned on the visible state of the local network rather than on private information alone.
}

{
Agents can also make a declaration when $f$ is true, to their neighbours, stating if and who they are going to attack. In this context, the agents may receive an extra section of the prompt stating that they are allowed to lie. 
}

{
The declaration mechanism is designed to make deception observable at the behavioural level. Rather than inferring deception from internal reasoning traces, we compare what an agent publicly declares with what it subsequently does. This follows recent approaches that define deception as a discrepancy between communicated intentions and later private actions \citep{taylor2025spontaneous}. The same distinction between public statements and private actions is central to work on strategic deception under pressure and long-horizon agentic interaction \citep{scheurer2024strategically,xu2026lhdeception}. In our setting, a false declaration occurs when an agent announces an intended attack target, or the absence of an attack, and then takes an inconsistent action in the following turn. Deception could be explicitly informed to agents when $d$ is true.
}

{
The reputation-memory condition extends this design by allowing agents to observe a compact history of their neighbours' truthfulness, if $m$ is true. This implements a minimal form of social memory: agents are not only exposed to current resource states and declarations, but also to whether previous declarations were reliable. This choice is motivated by work showing that reputation mechanisms can reduce commons-like failures in LLM multi-agent systems \citep{ren2025reputation} and by social-deduction simulations in which trust and deception unfold through repeated interaction \citep{curvo2025traitors}. The resulting experimental design therefore could separate three mechanisms: the availability of declarations, the explicit permission to lie, and the availability of reputation memory.
}

{
\begin{table}[ht]
  \centering
  \begin{tabular}{@{}p{4cm} p{10cm}@{}}
    \toprule
    \textbf{Feature} & \textbf{Effect when active} \\
    \midrule
    Neighbour information
      & Each agent observes the block holdings of its neighbours at the start of the turn, so production and attack decisions can be conditioned on the visible state of the local network rather than on private information alone. \\
    \addlinespace
    Future declarations
      & Each agent may declare its intended next-turn attack, and is shown both its own previous declaration and the declarations issued by its neighbours. \\
    \addlinespace
    Deception
      & Only in combination with future declarations: an agent is explicitly informed (through the prompting) that it is allowed to lie, so the talk channel can carry misleading as well as truthful signals. \\
    \addlinespace
    Reputation memory
      & Each agent is shown a summary of the past truthfulness of each neighbour, i.e.\ the historical consistency between what a neighbour declared and what it subsequently did, allowing reputation to be accumulated and used to evaluate other agents' declarations. \\
    \bottomrule
  \end{tabular}
  \caption{Prompt features given to LLM and their effect description.}
  \label{tab:features}
\end{table}
}

\subsection{Experimental Design}

\subsubsection{Fixed experimental condition}

{
All simulations share a common baseline configuration. Each run initializes $20$ agents on an Erd\H{o}s--R\'enyi network with target average degree $5$, evolving for at most $40$ time steps. A run terminates early if the shared biosphere is exhausted or only one agent remains. Every agent begins with $k_0 = 5$, $r_0 = 5$, and $g_0 = 10$. All LLM agents are driven by \texttt{gpt-5-mini-2025-08-07} trought OpenAI API. The initial biosphere level $b_0$ and the four binary communication features (neighbour information, future declarations, deception, and reputation memory) are the only parameters that vary across experimental conditions, together to the possibility to receive information regarding $b$ at each time step. 
}

{
\begin{table}
    \centering
    \begin{tabular}{cc}
        \hline
        Parameter & Value used \\
        \hline
        Number of agents & $20$ \\
        Network topology & Erd\H{o}s--R\'enyi \\
        Target average degree & $5$ \\
        Maximum number of time steps & $40$ \\
        Initial industrial capacity, $k_0$ & $5$ \\
        Initial military force, $r_0$ & $5$ \\
        Initial green capacity, $g_0$ & $10$ \\
        LLM model & \texttt{gpt-5-mini} \\
        \hline
    \end{tabular}
    \caption{Baseline simulation configuration with experimental parameters.}
    \label{tab:baseline_configuration}
\end{table}
}

\subsubsection{Explorative experiment}
{
The experiment crossed the four features (see Tab \ref{tab:features}) in a full combinatorial design. The neighbour-information feature $in$ was varied across two machines, taking $in = 0$ on one machine and $in = 1$ on the other, while the three remaining boolean features (future declarations $f$, deception $d$, and reputation memory $m$) were each varied on both machines. This yields $2 \times 2 \times 2 = 8$ communication regimes per machine, each combined with $6$ $b$ levels, for $48$ parameter combinations per machine and $96$ combinations in total.
We used six initial biosphere levels $b_0 \in \{1000, 2000, 4000, 6000, 8000, 10000\}.$
Since the LLM is the main source of stochasticity in this experiment, and given that each LLM run is relatively expensive, for these were preliminary explorations of the feature settings space, a single simulation was executed per parameter combination. 
We note that the deception feature is only behaviourally meaningful when future declarations are active. Obviously, this is also true for the memory-reputation features. The combinations in which deception is enabled without a declaration were retained for design symmetry but coincide with the corresponding no-deception runs. 
All other model parameters were kept fixed at their baseline values, so that the resulting design combines a full combinatorial exploration of the communication regime with a controlled sweep over the initial abundance of the shared environmental resource.
}

{
The final-state variables recorded include the number of completed time steps $T_f$, the number of surviving and eliminated agents (respectively $S_a$ and $E_a$), the total number of attacks $N_a$, the total number of production decisions $d_x$, where $x$ is the specific production, and the final quantities of brown, black, red, and green blocks, respectively $b_{t_f}$, $k_{t_f}$, $r_{t_f}$, and $g_{t_f}$. In addition, the time series of the four block types $\{(b_t, k_t, g_t, r_t)\}_{t=0}^{T_f}$ were stored for every run.
}

\subsubsection{Narrow space experiment}

The combinatorial sweep described above served as a preliminary exploration of the feature space, intended to locate the regions in which the LLM-driven dynamics are most informative. It revealed that a large part of this space is dominated by early extinction events: under the more severe scarcity conditions as well as when information on neighbours was off the biosphere collapsed  before the time limits was reached. On the basis of these observations, we concentrated the successive round of simulations on a narrower region of the parameter space that we found more interesting, namely one in which extinction events are comparatively rare. 

Specifically, this narrower design retains only the three highest initial biosphere levels,
\[
  b_0 \in \{6000, 8000, 10000\},
\]
with $in=1$ at all times, so that attention can be focused on the declaration, deception, and reputation channels. With future declarations $f = 1$ active, the two remaining features are fully crossed in a $2 \times 2$ design, with deception $d \in \{0,1\}$ and reputation memory $m \in \{0,1\}$; the two deception settings were again split across the two machines. 

In addition, we explore the regime in which future declarations are disabled altogether, so that deception and reputation memory are inactive as well. This yields five communication regimes, each evaluated across the three biosphere levels, as shown in Table \ref{tab:regimes}.

\begin{table}[ht]
  \centering
  \begin{tabular}{@{}c c c c c@{}}
    \toprule
    \textbf{Regime} & \textsc{Neighbours Information} & \textsc{Future declaration} & \textsc{Deception} & \textsc{Reputation memory} \\
    \midrule
    R1 & \textsc{true}  & \textsc{true}  & \textsc{false} & \textsc{false} \\
    R2 & \textsc{true} & \textsc{true}  & \textsc{false} & \textsc{true}  \\
    R3 & \textsc{true} & \textsc{true}  & \textsc{true}  & \textsc{false} \\
    R4 & \textsc{true} & \textsc{true}  & \textsc{true}  & \textsc{true}  \\
    R5 & \textsc{true} & \textsc{false} & ---            & ---            \\
    \bottomrule
  \end{tabular}
  \caption{The five communication regimes in the narrowed design. Regimes R1--R4 form a $2\times2$ crossing of deception and reputation memory under active future declarations; R5 disables future declarations, leaving deception and reputation inactive. Throughout, neighbour information is on ($\textsc{info\_neighbours}=\textsc{true}$), and each regime is evaluated at the three highest initial biosphere levels,  $b_0 \in \{6000, 8000, 10000\}$.}
  \label{tab:regimes}
\end{table}

\subsubsection{Biosphere level disclosure experiment}

The disclosure experiment is designed as another gaslighting experiment, specifically to check whether providing information about the global state of the biosphere $b$ can influence the agent's behaviour. 
We add a snippet to the prompt that additionally reports the current global brown level, ``\texttt{There are Br=$\langle$value$\rangle$
        left in the shared biosphere}'',
updated at every step, so that each agent $A_i$ observes how close the shared resource is to exhaustion and can condition its production on that trajectory.

Apart from this one line, the two conditions are identical to the previous experiments. So, the manipulation therefore isolates the effect of observing the shared resource from the effect of being told, abstractly, that it exists.

It keeps the gaslighting baseline, and the new disclosure runs directly comparable by holding the communication regime fixed: all four communication features ($ni$, $f$, $d$, and $m$) are active, so that biosphere visibility is the only feature that varies. We run both conditions at two initial biosphere levels, a low one and a high one,
\[
  b_0 \in \{2000, 10000\},
\]

\subsubsection{Rule-based experiments}

In the rule-based experiment, we investigated the same parameter space as in the LLM experiments in order to construct an interpretable reference model for comparison. The purpose of this experiment was to separate the dynamics generated by the sustainability game itself from the additional behavioural flexibility introduced by LLM-based decision-making. Rule-based agents follow explicit production and attack rules, so their behaviour can be directly linked to the parameters that define their preferences. They therefore provide a transparent benchmark against which the behaviour of LLM agents can be interpreted.

The rule-based and LLM experiments were aligned at the level of the simulation environment. In both cases, agents faced the same game structure, initial resource composition, and starting number of brown blocks. However, the behavioural parameters of the two agent types cannot be matched directly. The LLM agents make decisions through prompt-based natural-language reasoning, whereas the rule-based agents act according to explicit production and attack preference genes. For this reason, the rule-based production and attack genes were sampled independently and randomly, rather than being designed in advance to reproduce specific LLM behavioural tendencies.

Within each rule-based simulation run, all agents were assigned the same production and attack preference vectors, $p_i$ and $z_i$. This removes between-agent heterogeneity and makes each run interpretable as the outcome of a single population-level behavioural genotype. Differences between rule-based simulations can therefore be attributed to differences in the sampled behavioural rule, rather than to variation among agents within the same run.

This rule-based comparison is also consistent with the broader logic of agent-based modelling, where aggregate outcomes are explained by specifying explicit micro-level behavioural rules and observing the system-level dynamics they generate \cite{epstein1999agent,bruch2015agent}. In the present case, the rule-based model does not aim to reproduce the internal reasoning of LLM agents. Instead, it provides a controlled and explainable reference point for assessing how far the aggregate behaviour of LLM-based agents can be related to simpler behavioural tendencies.

\section{Results}

\subsection{The Effects of neighbour Information}
\begin{figure}[t]
    \centering
    \includegraphics[width=\linewidth]{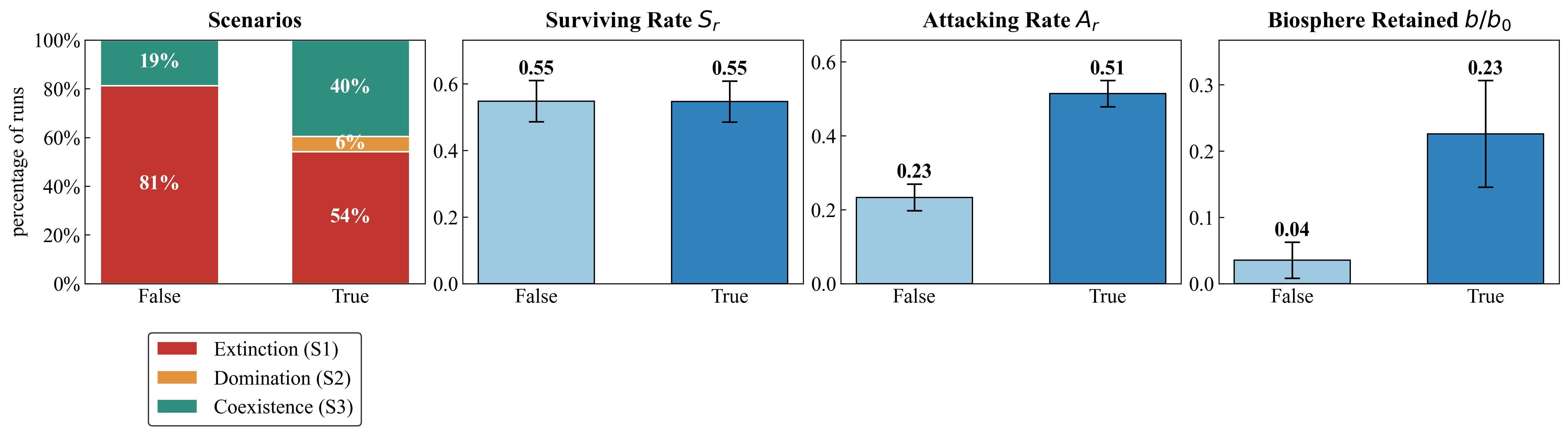}
    \caption{\textbf{Effect of neighbour information on system dynamics.} Results are pooled across all 96 scoping simulations (20 agents) and all combinations of initial resource levels and experimental parameters. Panels compare simulations without (\texttt{False}) and with (\texttt{True}) neighbour information. From left to right, the four panels show terminal outcome frequencies ($S1$ extinction, $S2$ domination, $S3$ coexistence), mean survival rate $E[S_r]$, total attacks $A_r$, and biosphere retention ($b/b_0$).}
    \label{fig:info_neighbours}
\end{figure}

We first ran a 96-simulation batch as described in the explorative experiment. From the results, we found that the possibility of accessing information from the neighbors ($I$ = True) is a major determinant of system behaviour and guided the design of our subsequent experiments (Fig.~\ref{fig:info_neighbours}).

Allowing agents to observe neighbours' resource stocks leaves the average survival rate unchanged ($S_r = 0.55$ in both conditions) but substantially alters the system's dynamics. With neighbour information $ni$ active, attacks increase almost threefold, biosphere retention rises from 3.5\% to 22.6\%, coexistence outcomes roughly double, and simulations last longer. This pattern holds across all resource levels: coexistence is consistently higher when neighbour information is available, reaching 1.00 versus 0.62 at ($b_0=10000$). Without neighbour information, agents act largely blindly. Attacks are rare, and most runs end in near-universal extinction (39 of 48 runs), leaving few brown blocks remaining. By contrast, visibility enables targeted behaviour and produces a richer outcome structure, including all three terminal regimes.

We therefore restrict the main study to the case where the information about the neighbors is active ($in = 1$), since the opposite regime is dynamically limited, with low attack rates, minimal biosphere retention, and predominantly extinction outcomes. More importantly, declarations, deception, and reputation only have meaning when agents can observe and reason about specific neighbours. 

\subsection{The Effects of declarations}
\begin{figure}[t]
  \centering
  \includegraphics[width=\linewidth]{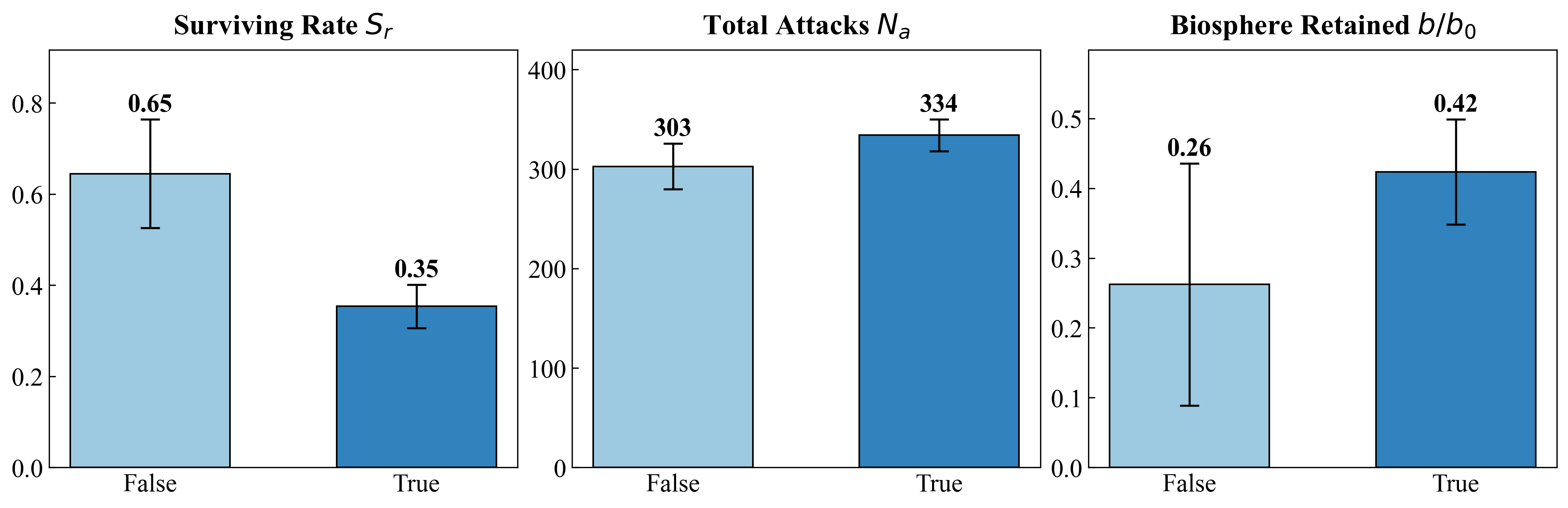}
  \caption{\textbf{Effect of future declarations on system dynamics.} Comparison of runs with future declarations disabled (R5) and enabled (R1--R4), with neighbors information active $I$ = True. Enabling declarations lowers extinction and raises coexistence and biosphere retention, while the per-agent survival rate falls and the total number of attacks rises slightly. Bars show means with error bars across runs.}
  \label{fig:declarations}
\end{figure}

\begin{figure}[t]
    \centering
    \includegraphics[width=\linewidth]{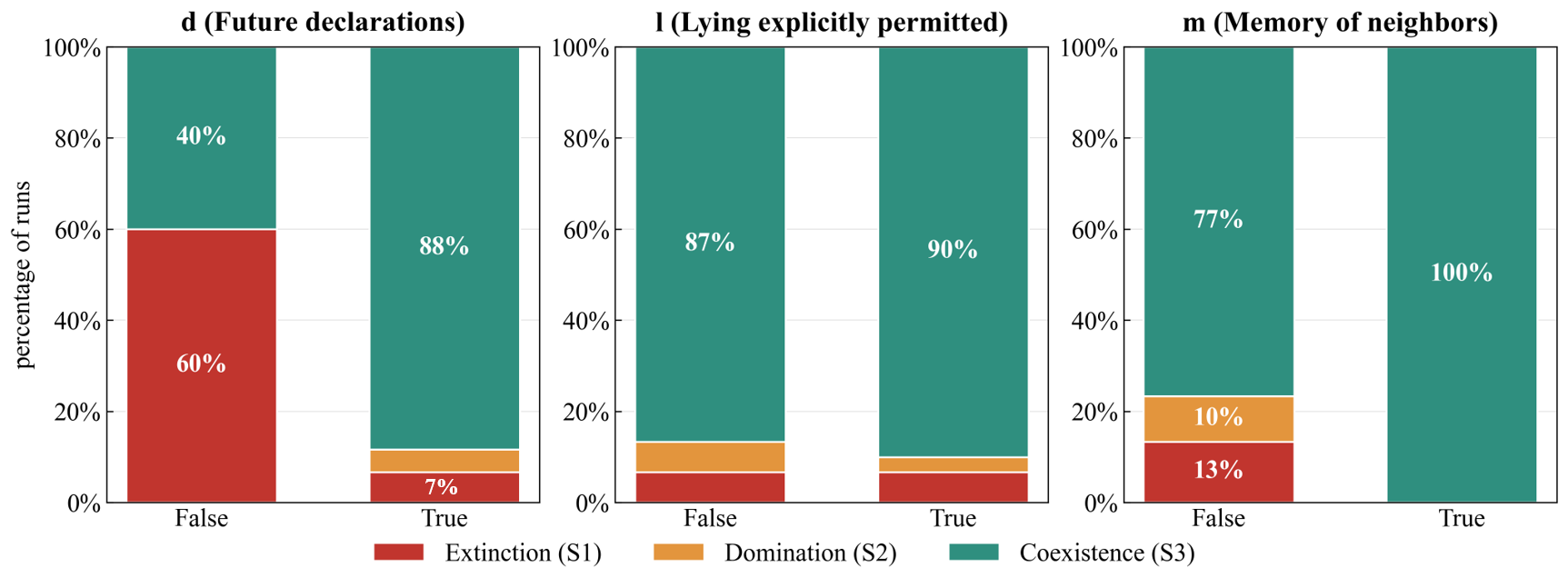}
    \caption{\textbf
    {Distribution of terminal outcomes under different communication settings.}  Results are shown for the main simulation batch with neighbour information enabled $(I=\mathrm{True})$ and $(b_0 \in {6000,8000,10000})$. Bars show the proportion of runs ending in extinction $(S_1)$, domination $(S_2)$, and coexistence $(S_3)$. The three panels compare the effects of future declarations, deception, and reputation memory, respectively. Future declarations and reputation memory shift the distribution towards coexistence, whereas deception alone has only a limited effect on terminal outcomes.}\label{fig:outcome-by-switch}
\end{figure}

With neighbour information fixed on, we next isolate the cheap-talk channel itself by comparing the no-declaration regime (R5) with the regimes in which future declarations are active (R1--R4). Opening the declaration channel changes the macro-dynamics markedly (Fig.~\ref{fig:declarations}). The fraction of runs ending in extinction falls from $40\%$ without declarations to $7\%$ with them, while coexistence rises from $60\%$ to $88\%$ and domination appears in a small share of runs ($5\%$), so that the population is far more likely to survive to the end of the horizon (Fig.~\ref{fig:outcome-by-switch}). The biosphere retained at the final step, $b_{t_f}$, increases accordingly from $26.2\%$ to $42.3\%$ of the initial brown stock, $b_0$. This gain in collective survival is not bought with peace: the per-agent survival rate drops sharply, from $0.65$ to $0.35$, and the total number of attacks edges up rather than down, from $303$ to $334$ per run.

Allowing agents to announce their intentions, therefore, does not pacify the population; it organises the conflict, concentrating attacks in a way that eliminates many individual agents while leaving a surviving core on a substantially better-preserved biosphere. The declaration channel thus behaves less as a coordination device that suppresses conflict and more as one that restructures it.

\subsection{The Effects of explicitly permitting lying and reputation memory} 
\begin{figure}[t]
  \centering
  \includegraphics[width=\linewidth]{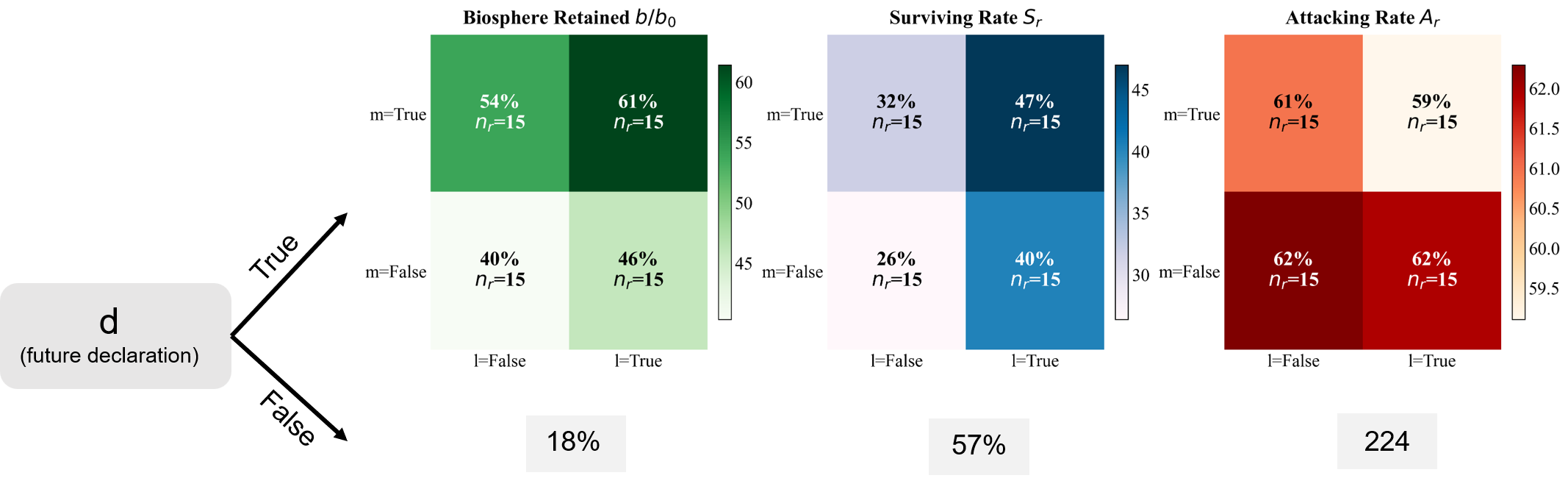}
  \caption{\textbf{Effect of deception and reputation memory on system dynamics.} Communication regime within $I$ = True. Rows indicate whether deception is disabled or enabled, and columns indicate whether reputation memory is disabled or enabled. The three panels show normalized biosphere retention $(b/b_0)$, mean survival rate $(E[S_r])$, and total attacks $(A_r))$. Deception and reputation memory jointly improve biosphere retention and survival, whereas attacks increase primarily with deception. Cell labels report the mean value and the number of simulation runs $n_r$.}
  \label{fig:lying-memory}
\end{figure}

Within the regimes that allow declarations, we cross the two remaining features in a $2\times2$ design: deception $d$ and reputation memory $m$. Both features push the system towards survival rather than collapse, and they do so roughly additively (Fig.~\ref{fig:lying-memory}). Enabling deception raises the biosphere retained from $40\%$ to $46\%$ without reputation memory and from $54\%$ to $61\%$ with it, and raises the survival rate from $26\%$ to $40\%$ and from $32\%$ to $47\%$, respectively. Reputation memory has a comparable effect along the other axis, lifting the biosphere retained from $40\%$ to $54\%$ when lying is off and from $46\%$ to $61\%$ when it is on. The most sustainable regime is therefore the one in which both features are active (R4): there, the biosphere retained reaches $61\%$ and the survival rate $47\%$, the highest of the design, and at the marginal level the introduction of reputation memory removes extinction almost entirely (the share of extinction runs $S1$ falls from $13\%$ to $0\%$ and coexistence rises to $100\%$) (Fig.~\ref{fig:outcome-by-switch}). As with declarations, the improvement in sustainability coincides with, rather than trades against, a high level of conflict: deception carries the largest attack counts of the whole design (around $344$--$345$ attacks per run, against $312$--$321$ without it), whereas reputation memory barely changes the amount of fighting. In this setting, then, an explicit licence to misrepresent intentions combined with the ability to track who has been truthful supports a more durable, even if more belligerent, equilibrium rather than destabilising it.

\subsection{Emergence and Forms of Deception}
Across declaration-enabled runs, we matched every stated next-turn intention against the agent's actual action in the subsequent turn. Deception emerged spontaneously rather than being explicitly permitted. Even when agents were not permitted to lie ($l = 0$), a substantial share of declarations were already dishonest, with a lie rate of 44\%. 

Allowing deception explicitly ($l = 1$), further increased these rates to 65\%. Therefore, permission amplifies an existing tendency toward deception rather than creating it. This pattern is large, consistent across initial biosphere levels ($b_0 = {6000, 8000, 10000}$), and robust across replications.

\begin{figure}[t]
    \centering
    \includegraphics[width=\linewidth]{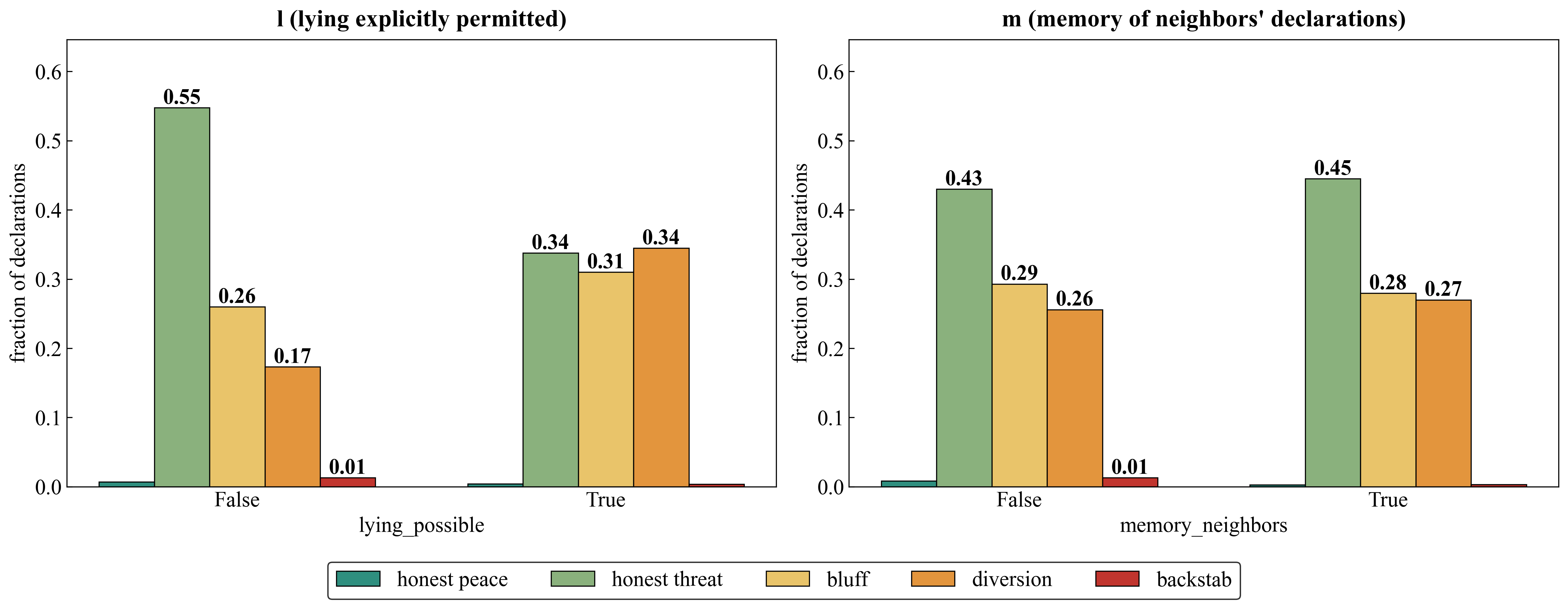}
    \caption{\textbf{How agents lie across different communication settings}
    (with declarations and neighbors information enabled; $N\approx 18{,}700$ declaration--action pairs). Each panel shows the composition of the five
    declaration types --- honest peace, honest threat, bluff, diversion, backstab --- for the
    False vs.\ True value of one switch. \textbf{Left:} permitting lying mainly inflates
    \emph{diversion} (announce one target, strike another) and \emph{bluffing}, while
    \emph{backstab} (promise peace, then attack) stays near zero. \textbf{Right:} reputation
    memory leaves the overall lie rate unchanged but selectively suppresses backstabbing.
    Deception is dominated by evasive and misdirecting statements rather than outright betrayal.}
    \label{fig:declaration-types}
\end{figure}

We categorized the declarations into honest peace (keeping a stated intent not to attack), honest threat (attacked exactly the announced target), bluff (declaring an attack and then not attacking), diversion (declaring one target while attacking another), and backstab (explicitly signalling peaceful intent and then attacking) (Table~\ref {tab:lie-taxonomy}). The structure of deception is equally notable (Fig.~\ref{fig:declaration-types}). Dishonesty is dominated by bluffing and diversion. However, backstabbing remains rare under all conditions, accounting for only about 0.4\%--2\% of declarations. The increase in deception produced by explicitly permitting lying ($l = 1$) operates almost entirely through diversion, which doubles (from 17\% to 34\%), while bluffing rises more modestly. In contrast, backstabbing shows little response to the permission treatment. Memory of neighbors ($m = 1$) leaves overall deception rates largely unchanged. Taken together, these results suggest that LLM agents may misrepresent their intentions even without explicit incentives to deceive. However, they tend to use evasive or misleading communication rather than directly breaking their stated peaceful commitments. Reputational memory further reduces the likelihood of such direct betrayals.

\begin{table}[t]
    \centering
    \caption{\textbf{Taxonomy of declarations}. Each next-turn intent declared at turn $t$ is compared with the agent's actual attack at turn $t{+}1$. ``Declared attack'' = a target was named; ``Attacked'' = an attack actually occurred.
    The three lie types differ in how they misrepresent the action.}
    \label{tab:lie-taxonomy}
    \begin{tabular}{@{}llccl@{}}
        \toprule
        \textbf{Category} & \textbf{Declared} & \textbf{Attacked} & \textbf{Lie} & \textbf{Description} \\
        \midrule
        Honest peace   & no attack   & no attack       & no & kept a stated intent not to attack \\
        Honest threat  & attack agent 1  & attack agent 1   & no & attacked exactly the announced target \\
        \addlinespace
        Bluff          & attack agent 1  & no attack        & yes & threatened, then did not attack \\
        Diversion      & attack agent 1  & attack agent 2   & yes & announced one target, struck another \\
        Backstab       & no attack   & attack agent 1       & yes & promised peace, then attacked \\
        \bottomrule
    \end{tabular}
\end{table}

\subsection{Comparison between rule-based agents and LLM-based agents}

\begin{figure}[t]
  \centering
  \includegraphics[width=\linewidth]{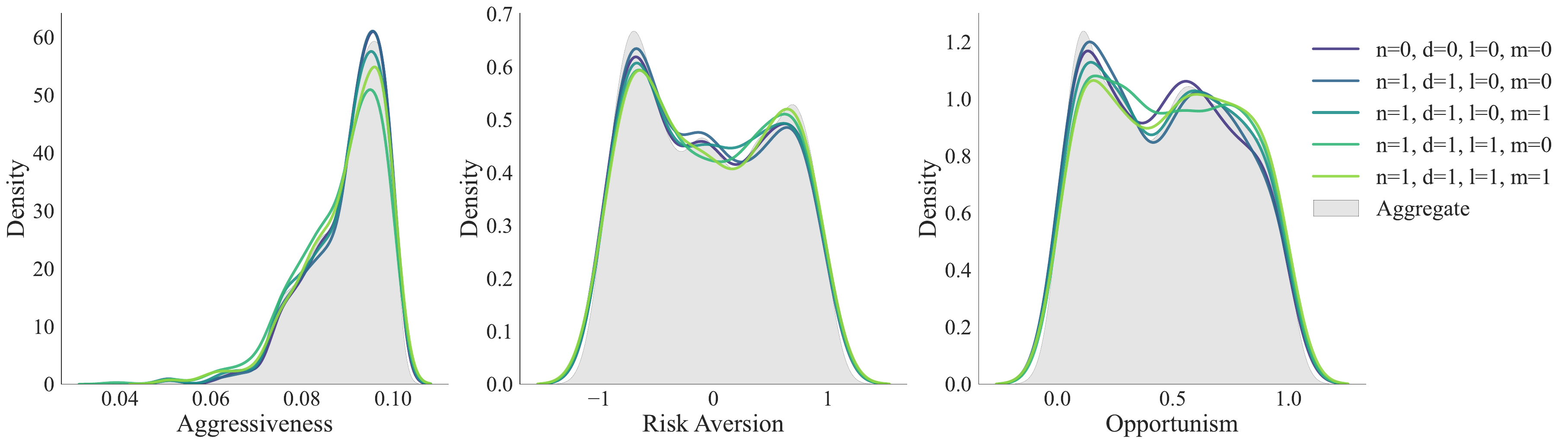}
  \caption{\textbf{Macroscopic trends in LLM population behavior can be reproduced by rule-based populations, while effects of specific prompt configurations cannot be captured.} For each LLM experiment, the top 1$\%$ best-matching rule-based simulations were identified by minimizing a combined similarity measure based on the mean absolute percentage error (MAPE) across resource time-series (browns, blacks, greens, and reds) together with the difference in attack rate per agent per timestep between LLM and rule-based simulations, $r_a$. Similarity was quantified through the Euclidean distance to the ideal point (($\mathrm{MAPE}=0,\Delta r_a=0$)). From left to right panels show the distributions of aggressiveness, risk aversion, and opportunism among the top matching rule-based simulations across all LLM experimental conditions for an initial brown population ($b_0 = 10000$). Each colored curve corresponds to a different LLM prompt setup, as specified in the legend, while the light-gray shaded region represents the aggregate distribution across all experimental conditions. 
}
\label{fig:gene_distribution}
\end{figure}

In Figure \ref{fig:gene_distribution}, panels show the distributions of aggressiveness, risk aversion, and opportunism (respectively, the genes $z_{opp}$, $z_{risk}$, and $z_{ogg}$ among the top matching rule-based simulations across all LLM experimental conditions for an initial brown population ($b_0 = 10000$). Each colored curve corresponds to a different LLM prompt setup, as specified in the legend, while the light-gray shaded region represents the aggregate distribution across all experimental conditions. Although the best-matching rule-based simulations are not randomly distributed across the sampled behavioural space, but instead concentrate around specific values of aggressiveness, risk aversion, and opportunism, this match should be interpreted with caution. The rule-based agents can reproduce some aggregate LLM outcomes, such as resource trajectories and attack rates, but they do not necessarily reproduce the way LLM behaviour changes across different prompt configurations. Thus, matching macroscopic dynamics does not imply that the rule-based model captures the condition-specific effects of neighbour information, declarations, deception, or reputation memory.

\subsection{Disclosing the global biosphere}
In all previous experiments, agents could only observe their neighbours and did not know the global state of brown stocks. Also, agents believe they can repair the environment with green-to-brown actions, but they actually cannot. We test whether telling agents the true global brown level changes their behaviour while keeping the same (limited-resource) dynamics, at two initial resource levels ($b_0\in\{2000, 10000\}$, all switches on, five replicative experiments each). 

\begin{figure}[t]
    \centering
    \includegraphics[width=\linewidth]{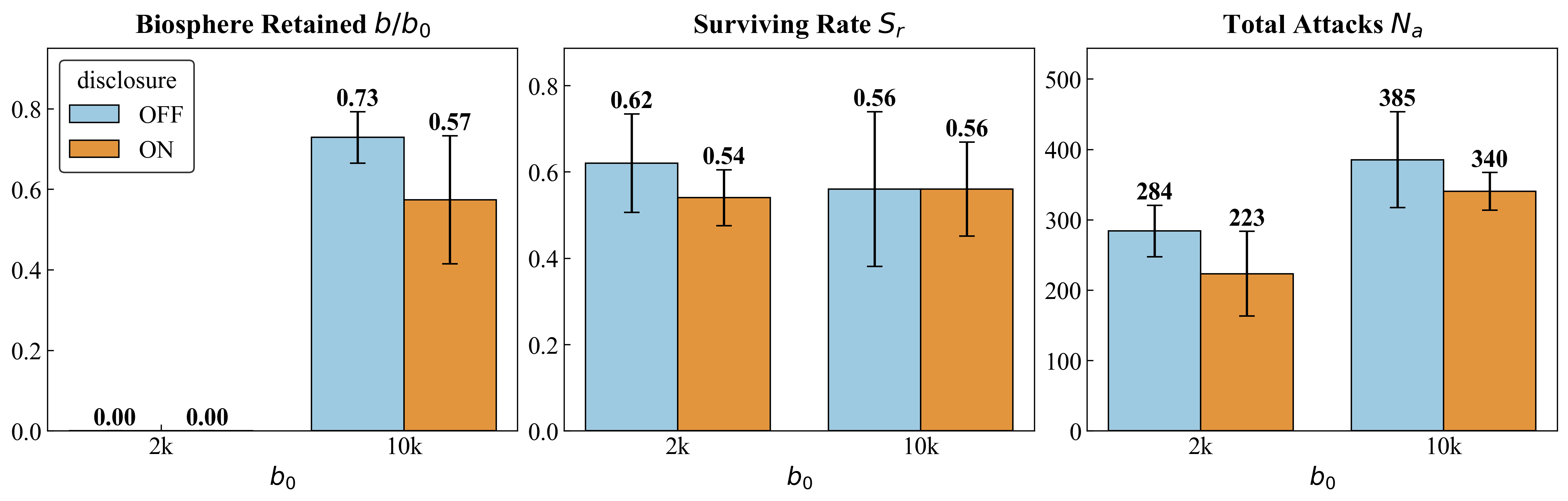}
    \caption{\textbf{Effect of disclosing the global biosphere stock on system behaviour} (all switches on; $b_0\in\{2000,10000\}$, 5 replications each; disclosure OFF vs ON). From left to right, the three panels show biosphere retained (\% of $b_0$), survival rate ($S$), and total attacks ($A$). Disclosure leaves the terminal regime and survival unchanged, but reduces biosphere retention at $b_0=10000$ and lowers attacks at $b_0=2000$. Bars are means; error bars are $95\%$ CIs.}
    \label{fig:disclosure-macro}
\end{figure}

We find disclosure does not change the terminal scenarios, which are highly constrained by the limited initial resource. Every $b_0=2000$ run still ends in extinction, and every $b_0=10000$ run ends in coexistence, regardless of disclosure, and survival is essentially unchanged (Figure~\ref{fig:disclosure-macro}). What disclosure does affect is environmental restraint and the volume of conflict. At the rich level ($b_0=10000$), the fraction of biosphere retained falls markedly when agents are informed (from $73\%$ to $57\%$), and at the scarce level $b_0=2000$, total attacks drop (284 to 223). In other words, once agents are told the stock is safe, they consume it less cautiously, while knowing the stock is doomed dampens fighting, but neither shifts the final outcome.

\begin{figure}[t]
    \centering
    \includegraphics[width=\linewidth]{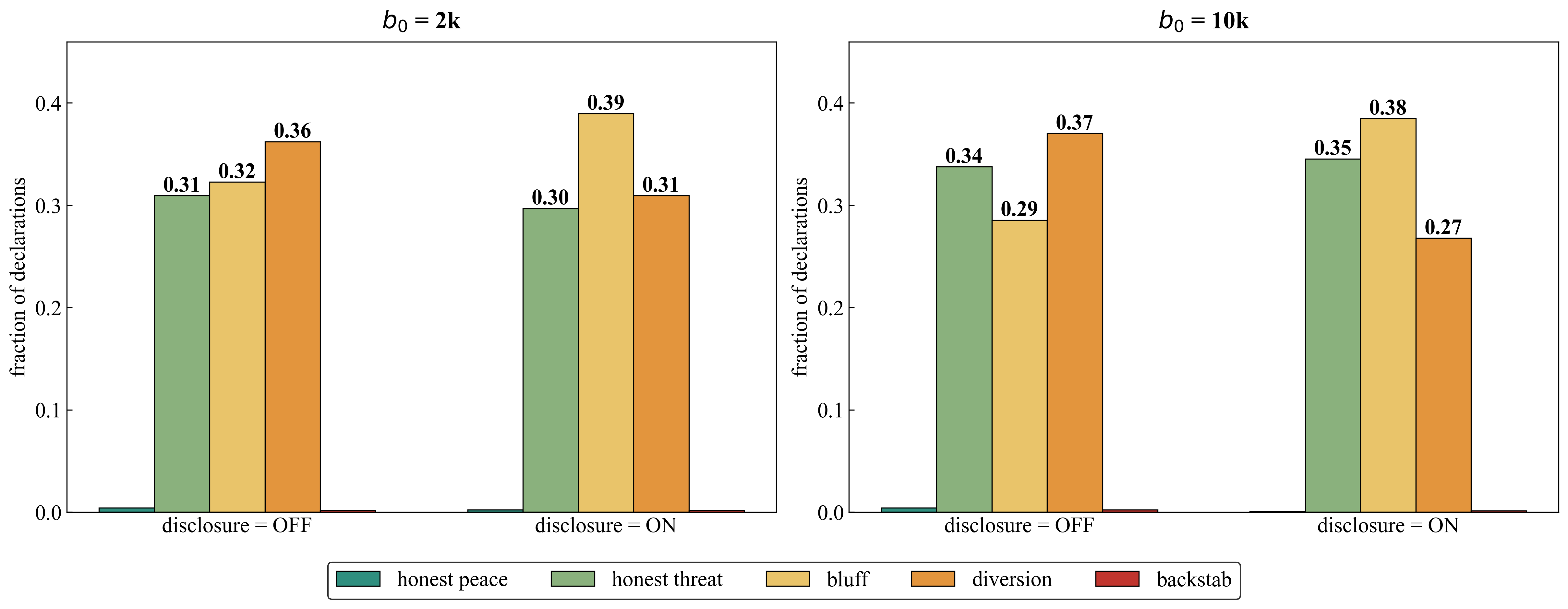}
    \caption{\textbf{Disclosure shifts the form of deception.}
    Composition of the five declaration types under disclosure OFF vs ON, for each $b_0$. The overall lie rate is unchanged, but diversion falls and bluffing rises in both resource regimes, while backstabbing stays near zero.}
    \label{fig:disclosure-lying}
\end{figure}

Disclosure also reshapes how agents lie without changing how often (Figure~\ref{fig:disclosure-lying}). The overall lie rate is essentially constant ($65\%$–$70\%$ across both conditions), but the composition shifts consistently between resource levels. Diversion declines while bluffing rises, and backstabbing remains negligible throughout. Access to global information thus softens deception from active misdirection toward empty threats, a robust effect independent of the resource regime. 

\section{Discussions}

\subsection{Emergence of Lying in LLM and the effect of Gaslighting}

{
Our results demonstrate that deceptive communication emerges naturally in LLM-based sustainability games, extending beyond simple information exchange to include lying behaviour \cite{riedl2026emergent}. Although declarations of future intentions improve collective sustainability by enabling agents to anticipate one another's actions, they also create opportunities for strategic manipulation of neighbouring agents \cite{wang2023avalon}. Rather than being an isolated behaviour, deceptive signalling becomes part of the decision-making process, allowing agents to influence others' expectations while pursuing their own objectives \cite{backmann2025ethics}. Nevertheless, it is interesting that LLM-agents are almost not performing the worst style of lying in this scenario, which is backstabbing. This could be a consequence of the fear of retaliation, deriving from pre-training experience with a game-theoretical framework, or a purely emergent property \cite{lynch2025agentic}; future research could investigate it more thoroughly.  
}

{
Moreover, the experiments further show that explicitly permitting deception amplifies these behaviours, whereas reputation mechanisms reduce their effectiveness by making past actions available for future decisions. These findings suggest that communication in LLM-agent societies is inherently dual-purpose: it supports coordination while simultaneously creating opportunities for strategic misinformation \cite{wang2023avalon}. Consequently, understanding agentic behaviour requires considering not only what information is exchanged but also whether that information can be trusted \cite{de2025llm}
}

\subsection{Relationship of Rule-based and LLM-based ABM}
{
Beyond the specific empirical results, the framework shows how traditional rule-based agents and LLM approaches can be used together \cite{taillandier2025integrating,giabbanelli2026guide}. The rule-based model provides a transparent and explainable structure \cite{bonabeau2002agent}, in our case with the environmental and resource dynamics: blocks are produced, converted, consumed, and depleted according to explicit rules. LLM-based agents, by contrast, introduce more flexible decision-making, because they can interpret natural-language information, respond to neighbour behaviour, make declarations, and adapt their actions across different prompt conditions \cite{gao2024large}. In this sense, the rule-based component provides a sort of explainable formalization to the sustainability problem as well as allowing to understand whether the LLM-behaviour is reducible to a set of algorithmic rules.
This combination preserves the interpretability of classical ABMs while allowing behaviours such as communication, deception, and reputation to emerge from agent interactions rather than being explicitly programmed. In contrast to previous sustainability-game frameworks based primarily on predefined behavioural rules \cite{bertolotti2024balancing}.
}

{
The hybrid design adopted in this study provides a transparent framework for distinguishing the influence of language-driven reasoning from the underlying sustainability dynamics. Rather than replacing conventional ABMs, LLM reasoning is compared within an established rule-based environment, allowing strategic communication to emerge while preserving the underlying ecological game of \cite{bertolotti2024balancing,taillandier2025integrating}. The comparison with rule-based agents further provides an interpretable behavioural reference for understanding how communication, deception, and reputation contribute to the patterns observed in the experiments \cite{giabbanelli2026guide}. By combining transparent environmental processes with flexible language-driven decisions, the framework bridges classical sustainability modelling and recent advances in LLM-based multi-agent systems \cite{bonabeau2002agent,epstein2012generative}.
}

\subsection{Limitations}
{
Although our framework provides new insights into communication, deception, and reputation in LLM-based sustainability games, several limitations remain, pointing to promising directions for future research. First, the proposed framework considers a simplified setting in which agents share identical initial conditions and interact through a fixed network structure. While useful, this is an obvious limitation, given that the results should be validated in different settings \cite{galan2009errors}. 
Second, LLM-agent behaviour is generated through prompting rather than learning, making the observed dynamics dependent on the underlying language model and prompting strategy. The prompt is not updated either intra- or inter-simulation. Since adaptive multi-agent systems are progressively more relevant, this is a limitation for policy-makers and practitioners. 
Third, agents follow fixed decision policies throughout each simulation, whereas real social systems continuously adapt their behaviour in response to changing conditions and the observed success of others. }


\section{Conclusions}

{
This paper examined whether lying can emerge among LLM-based agents in a competitive sustainability game with global deception. We implemented an ABM in which agents manage shared resources, compete through attacks, and communicate through declarations, deception, and reputation memory. The main results show that deception emerges even when agents are not explicitly allowed to lie, and that communication reshapes conflict rather than simply reducing it. Neighbour information, declarations, deception, and reputation memory can improve biosphere consumption and the rate of collective survival, though often accompanied by higher levels of conflict. Also, rule-based agents could serve as an interpretable baseline for understanding LLM-based behaviour.
}

{The findings may inform the design and governance of autonomous multi-agent systems operating under shared resource constraints. Beyond sustainability, similar dynamics arise in geopolitical coordination, where climate agreements, resource security, and strategic competition depend on credible commitments and trust among decentralized actors \cite{ostrom1990commons,levin2013social}. Our results suggest that information sharing alone may be insufficient to sustain collective action when incentives for strategic misrepresentation exist, whereas reputation and social memory can strengthen trust across repeated interactions. These insights are also relevant to emerging AI ecosystems, including collaborative AI assistants and decentralized autonomous organizations, where autonomous agents increasingly communicate and make decisions on behalf of users. Although the framework is intentionally simplified, it illustrates how communication, credibility, and accountability can shape collective outcomes in systems that depend on sustained cooperation.}

{
Future work will address the previously stated limitations by incorporating heterogeneous agent populations, dynamic interaction networks, richer communication protocols, and adaptive learning mechanisms. Another promising direction is to integrate evolutionary game-theoretic strategy updating, allowing agents to revise their behaviour based on the relative success of neighbouring strategies, while communication, deception, and reputation co-evolve over repeated interactions \cite{nowak2006five,szolnoki2007cooperation}. 
Finally, investigations on LLM agents' adaptation and prompt sensitivity of the results could result in a better understanding of both the system behaviour and LLM collective features.
}

\section{Acknowledgements}
This work is the output of the workshop Complexity72h by Complexity Next Gen, held at Northeastern University London, London, UK, 22-26 June 2026. www.complexitynextgen.org/complexity72h/.

\bibliographystyle{unsrtnat}
\bibliography{references}  






\end{document}